# A Wearable Medical Sensor for Provisional Healthcare

Amir Javadpour
Faculty of New Sciences and Technologies
University of Tehran
Tehran, Iran
Email: a.javadpour@ut.ac.ir

HamidrezaMemarzadeh-Tehran
Faculty of New Sciences and Technologies
University of Tehran
Tehran, Iran
Email: hmemar@ut.ac.ir

*Abstract*—Thispaper presents the design and realization of a context-aware wireless health monitoring system for recording the heartbeat (HR) and respiration (RR) rate based on an indirect measurement approach. The system consists of a contact-less medical sensor as well as a communication infrastructure for handling the transmission and reception of the measured results. The contact-less sensor includes a highly sensitive tri-axial accelerometer, an accurate temperature and air pressure sensor that enable one to inspect patients' health condition by continuously monitoring of two critical signs related to the cardiorespiratory system. The developed system can also be utilized in performing a number of long-term inspection on the heart and lungs while measuring the HR and RR values in addition to calculating the HR and RR ratio, which is denoted by HRR. The obtained results show the potential of the developed system for versatile monitoring applications applied to telemedicine.

## I. INTRODUCTION

The aging population and sedentary life style have caused a significant growth in the number of patients who are diagnosed for cardiovascular, respiratory system and diabetes diseases [l]. As a consequence, an enormous number of patients will increasingly need medical and health-related services annuallythat results in shortcoming of basic requirements. To alleviate the loads enforced on the conventional healthcare system, it is suggested that the patients who are diagnosed for one of the above-mentioned diseases be provided with a portable and light-weight health monitoring system in order to reduce the number of unnecessary referrals to clinics or caregiving organizations. Recently, the telemedicine, a systematic approach for revolutionizing the conventional healthcare system, has become the technology of choice in resolving the drawbacks associated with the conventional healthcare system in which the information technology (IT) as an enabling tool in conjunction with using low-power radio frequency (RF) electronics are employed [2]. The healthcare system benefiting the telemedicine technology can offer a variety of location-based services to patients such as remote vital signs monitoring, health condition inspection and pre-diagnosis at home or alternatively during daily living life [3]. As it was clearly noted, the continuous monitoring of the human vital signs as the key element in telemedicine can provide a powerful mean for obtaining long-term information about patient's health condition as opposed to conventional care-giving procedures which are usually available at hospitals orclinicswhile the patients are at rest position.

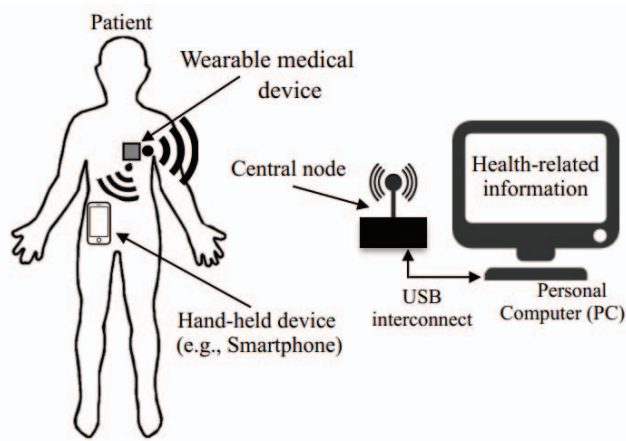

Fig. 1.Schematic of an example health monitoring system consisting of a wearable medical sensor.

Such the mobility and location-based characteristics become available.by incorporating the concept of telemedicine enabling the practitioners to obtain a more detailed information about the patients physiological behavior and general health condition. Usually, important vital signs such as heart rate, respiration and temperature, for example accounted for seizure, can be obtained by using individual sensors or medical devices such as electrocardiogram (ECG) [4], phonocardiogram (PCG) [5] and thermometers [6], which are bulky, heavy and uncomfortable and sometimes delegated to be carried by patients. Such the monitoring devices, which are mainly based on a direct measurement approach, besides not being portable are also sensitive to ambient and surrounding environment noise as well as the artifacts caused because of the movements of the patients and displacement of the contacting electrodes. Therefore, the obtained vital signs by using the conventional devices are considerably prone to errors. In order to overcome the drawbacks associated with the direct measurement approach in obtaining the vital signs, it was found that it is feasible to develop a monitoring system consisting a number of wireless medical sensors which are able to measure the human vital signs in an indirect approach. To this end, it is suggested to develop a low-profile and comfortably wearable medical sensor capable of measuring the human vital signs in an indirect fashion [7].In this paper, a design and realization of a vital signs monitoring system based on indirect measurement approach is presented.

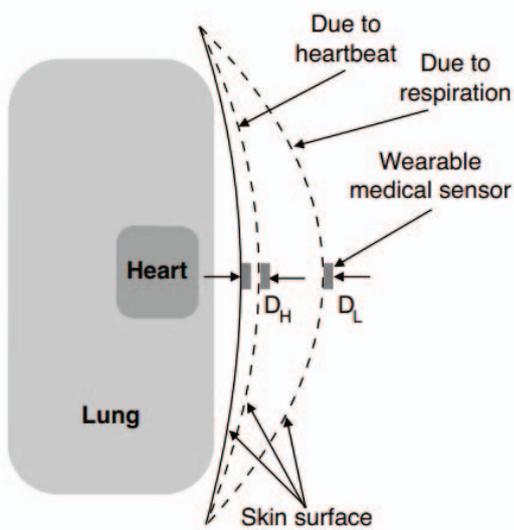

Fig. 2.Schematic depicting the mechanism used for measuring the heartbeat and respiration rate in an indirect fashion.

The monitoring system is able to measure both the heartbeat (HR) and respiration rate (RR)as well as the human body temperature (HBT) during daily activities without constraining the patient's movement. The medical sensor functionality for measuring the heartbeat and respiration rate is based on measuring two different movements which are due to the pumping of blood into the vessels and changing of the volume of the lungs during the inhalation and exhalation process.

## II. INDIRECT VITAL SIGN MEASUREMENT MECHANISM

As it is shown in Fig. 2, the functionality of the cardiorespiratory system causes small movements in the chest and abdomen area. The movements can be categorized into two distinct short and long-term deformations which are considered as indicators for heartbeat and respiration, respectively. The chest movement due to lungs activity denoted by $D_L$ is larger with slower frequency whilst a higher frequency with a short duration is expected as the result of heart's muscle activity $(D_H)$.Since the surface of chest or abdomen moves back and forth due to the lungs activity, the obtained movement is a superposition of both movements namely,$D_{Total}=C_1\ D_L + C_2\ D_H$ ,where$C_1$ and $C_2$ are constants. To this end, a medical sensor capable of recording both short and long duration movements of a patient's chest or abdomen can enable one to indirectly measure both heartbeat and respirate rate. Additionally, by incorporating an accurate temperature sensor, the skin temperature of the patient as an indicator for the intensity of activity can be obtained.

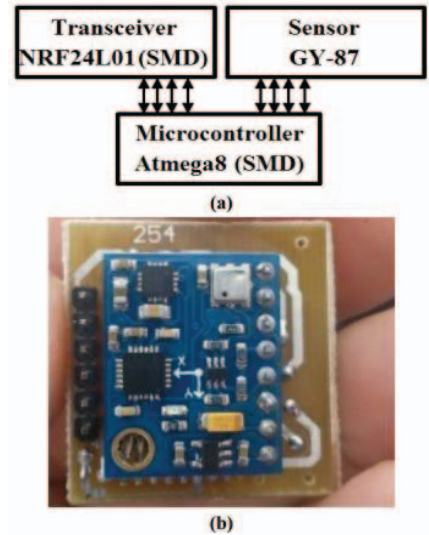

Fig. 3.The developed medical sensor; (a) schematic of internal components, (b) a photograph of the developed medical sensor.

## III. THE DEVELOPED MONITORING SYSTEM

The schematic of the monitoring system is represented in Fig. 1 which consists of a medical sensor and a central node for collecting the measurement data. In the followingof section III, the design and implementation of the medical sensor and the central node are described.

### A. The Medical Sensor

In order to measure the short and long-term movement $(D_{Total})$, one can use an accelerometer instead of positioning devices, which are not generally accurate. The $D_{Total}$can be calculated based on the measurement results obtained from the accelerometer by using (1),

$$D_{Total} = \iint_T a_{Total}\, dt \quad (1)$$

where$a_{Total}$denotes the total acceleration and $T$indicates the measurement duration. The calculation of $a_{Total}$is based on the acceleration values along three x, y, z-axis which are denoted by$a_x, a_y$ and$a_z$respectively,

$$a_{Total} = \sqrt{a_x^2 + a_y^2 + a_z^2} \quad (2)$$

To this end, a medical sensor comprising three essential components namely a sensitive tri-axial accelerometer (GY-87,U-blox), microcontroller (Atmega8, Atmel) as well as a transceiver (nRF24, Nordic Semiconductor) for transmitting the measured signals to a remote device (i.e., central node) was developed. Fig. 3 represents the building blocks and a photograph of the developed medical device.

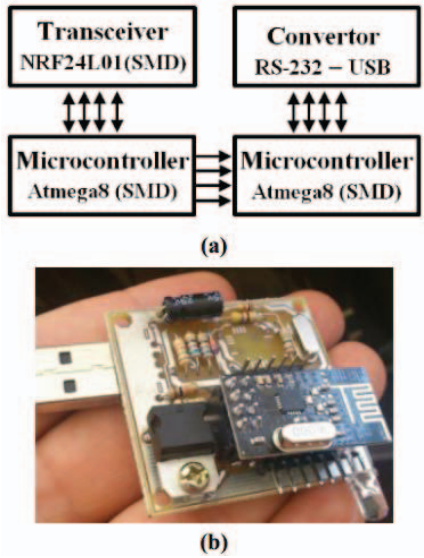

Fig. 4. The developed central node; (a) schematic of internal components, (b) a photograph of the developed central node.

### B. The Central Node

In the developed monitoring system, the central node, as shown in Fig. 4, is responsible for gathering and transferring the measured signals to a personal computer (PC) or a hand-held device (e.g., smartphone) was realized. The central node includes two identical microcontrollers (Atmega8, Atmel). One microcontroller is used to receive data from the transceiver and the other one performs as a RS-232 to USB port converter. The medical sensor and central node communicate at the operation frequency of 2.45 GHz, which is allocated at the ISM (Industrial, Scientific and Medical) band.

## IV. EXPERIMENTAL EVALUATION OF THE DEVELOPED MEDICAL DEVICE

In order to evaluate the performance of the developed medical sensor and the achieved sensitivity in measuring a volunteer's vital signs (i.e., heartbeat and respiration rate), the sensor was placed on a volunteer's clothing as shown in Fig. 5. In this experiment, the volunteer held the breath (RR→ 0) for almost 4.5 sec. so as to allow only the heartbeat to be measured by the medical sensor. Figure 6 shows the measured acceleration ($a_{Total}$) which corresponds to the volunteer's heartbeat. It can be also observed that the device is sensitive to pick-up short-term skin movement ($D_H$) even if it is placed on the volunteer's clothing. Another measurement when the medical device placed on the chest of a volunteer's chest, close to his heart, was performed.

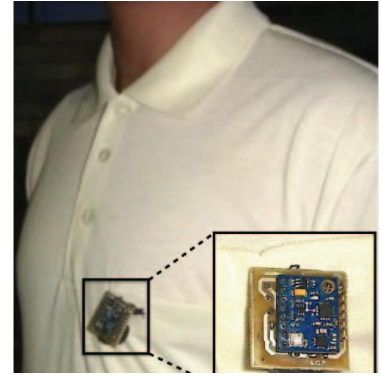

Fig. 5. The photograph showing the medical sensor worn by a volunteer.

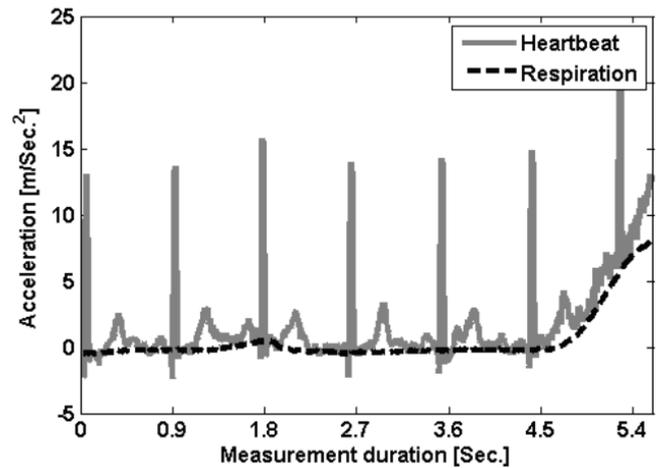

Fig. 6. The heartbeat of a volunteer obtained from measurement while holding the breath.

The measurement results, as represented in Fig. 7, reveal that the device is able to measure both the short and long-term movements due to heart and lungs activity ($D_{Total}$). In Fig. 7, the measurement results related to the volunteer who breathed normally for a period of 7.2 sec. is represented. As it can be seen, the HR and RR are 4 and 1, respectively. By using (3) one can obtain the ratio between the HR and RR as an indicator for ensuring a healthy cardiorespiratory system. The obtained value for HRR for this experiment is 4. For a healthy person, the HRR values are normally between 3 to 8. Figure 8 represents the heartbeat and respiration rate of the volunteer while running with HR=9 and RR=3.

$$HRR(T_{body}, T_{Air}) = \frac{HR}{RR} \quad (3)$$

It is noteworthy to mention that the medical sensor is equipped with a temperature and air pressure sensors to enable the system to be aware of the context of the patient namely, the intensity of the activity and the latitude (e.g., sea surface or mountain) at which a volunteer's health condition is monitored. As a further inspection on a patient's health condition, the developed medical device was set on the patient's clothing while measuring the heartbeat until the patient coughs, as the obtained results shown in Fig. 9, indicating approximately the time $(t = 5.85$ sec.) at which a cough took placed.

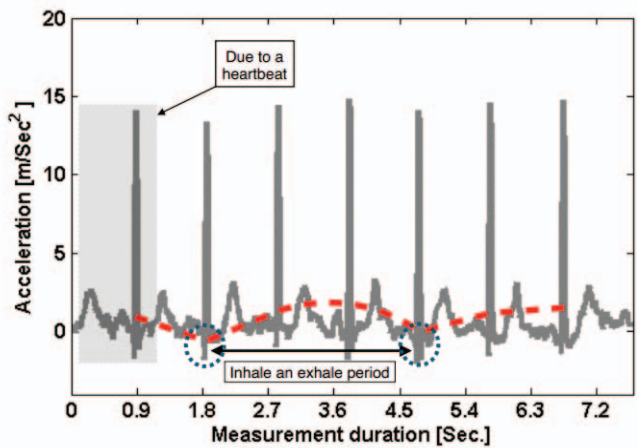

Fig. 7. Heartbeat and respiration rate of a volunteer obtained using the medical sensor.

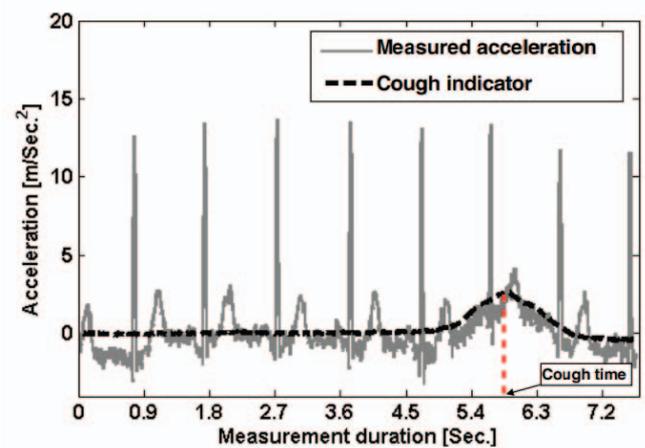

Fig. 9. The heartbeat and respiration rate of the volunteer while coughing.

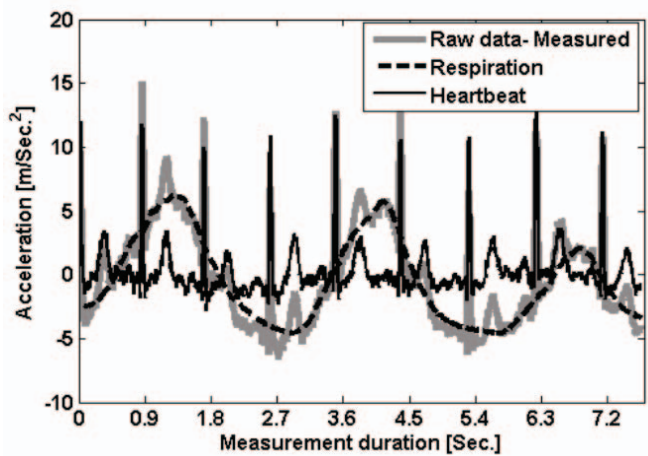

Fig. 8. The heartbeat and respiration rate of the volunteer while running.

## V. CONCLUSION

In this paper, the design and implementation of a health monitoring medical device for recording the heartbeat and respiration rate based on an indirect approach was described. Also, the system consists of a medical sensor as well as a central node which is used to transfer the measurement results to a PC. Also, the monitoring system can be used for monitoring the signs related to the patient's heart and lungs health such heartbeat, respiration as well as the duration of inhalation and exhalation.